\documentclass[aps,prl,twocolumn,nofootinbibs,tightenlines,superscriptaddress]{revtex4-2}

\usepackage[utf8]{inputenc}
\usepackage[T1]{fontenc}

\usepackage{lmodern}
\usepackage{amsmath,amsfonts,amssymb}
\usepackage{graphicx}
\usepackage{trimclip,stackengine}

\usepackage{lipsum}
\usepackage{comment}

\usepackage{hyperref}

\usepackage[usenames,dvipsnames]{xcolor}
\hypersetup{colorlinks=true, linkcolor=blue!50!black, urlcolor=blue!50!black, citecolor=blue!50!black}
\usepackage[all]{hypcap}
\usepackage{xcolor}
\usepackage{tikz}

\newcommand{\subref}[2]{\hyperref[#1]{#2}}

\usepackage[normalem]{ulem}

\newcommand\add[1]{\textcolor{blue}{#1}}

\usetikzlibrary{matrix}
\usetikzlibrary{calc,fit}
\tikzset{%
  highlight1/.style={rectangle,rounded corners,color=red!,fill=red!15,draw,fill opacity=0.5,thick,inner sep=0pt}
}
\tikzset{%
  highlight2/.style={rectangle,rounded corners,color=green!,fill=green!15,draw,fill opacity=0.5,thick,inner sep=0pt}
}

\begin{document}

\title{Dimensional crossover via confinement in the lattice Lorentz gas}



\author{Alessio Squarcini}
\email{alessio.squarcini@uni.lu}
\affiliation{Institut f\"ur Theoretische Physik, Universit\"at Innsbruck, Technikerstra{\ss}e~21A, A-6020 Innsbruck, Austria}

\affiliation{Complex Systems and Statistical Mechanics, Department of Physics and Materials Science, University of Luxembourg, 30 Avenue des Hauts-Fourneaux, L-4362 Esch-sur-Alzette, Luxembourg}

\author{Antonio Tinti}
\affiliation{Dipartimento di Ingegneria Meccanica e Aerospaziale, \\ Sapienza Universit\`a di Roma, via Eudossiana 18, 00184 Rome, Italy}

\affiliation{Laboratory of Molecular Simulation (LSMO), Institut des Sciences et Ingénierie Chimiques, École Polytechnique Fédérale de Lausanne (EPFL), 1950, Sion, Switzerland}

\author{Pierre Illien}
\affiliation{Sorbonne Universit\'e, CNRS, Laboratoire PHENIX (Physico-Chimie des Electrolytes et Nanosyst\`emes Interfaciaux), 4 Place Jussieu, 75005 Paris, France}

\author{Olivier B\'enichou}
\affiliation{Laboratoire de Physique Th{\'e}orique de la Mati{\`e}re Condens{\'e}e, CNRS/UPMC, 4 Place Jussieu, F-75005 Paris, France}

\author{Thomas Franosch}\email{thomas.franosch@uibk.ac.at}
\affiliation{Institut f\"ur Theoretische Physik, Universit\"at Innsbruck, Technikerstra{\ss}e~21A, A-6020 Innsbruck, Austria}

\date{\today}

\begin{abstract}
We consider a lattice model in which a tracer particle moves in the presence of randomly distributed immobile obstacles. The crowding effect due to the obstacles interplays with the quasi-confinement imposed by wrapping the lattice onto a cylinder. We compute the velocity autocorrelation function and show that already in equilibrium the system exhibits a dimensional crossover from two- to one-dimensional as time progresses. A pulling force is switched on and we characterize analytically the stationary state in terms of the stationary velocity and diffusion coefficient. Stochastic simulations are used to discuss the range of validity of the analytic results. Our calculation, exact to first order in the obstacle density, holds for arbitrarily large forces and confinement size.
\end{abstract}


\maketitle



\begin{figure}[htbp]
\centering
\includegraphics[width=\columnwidth]{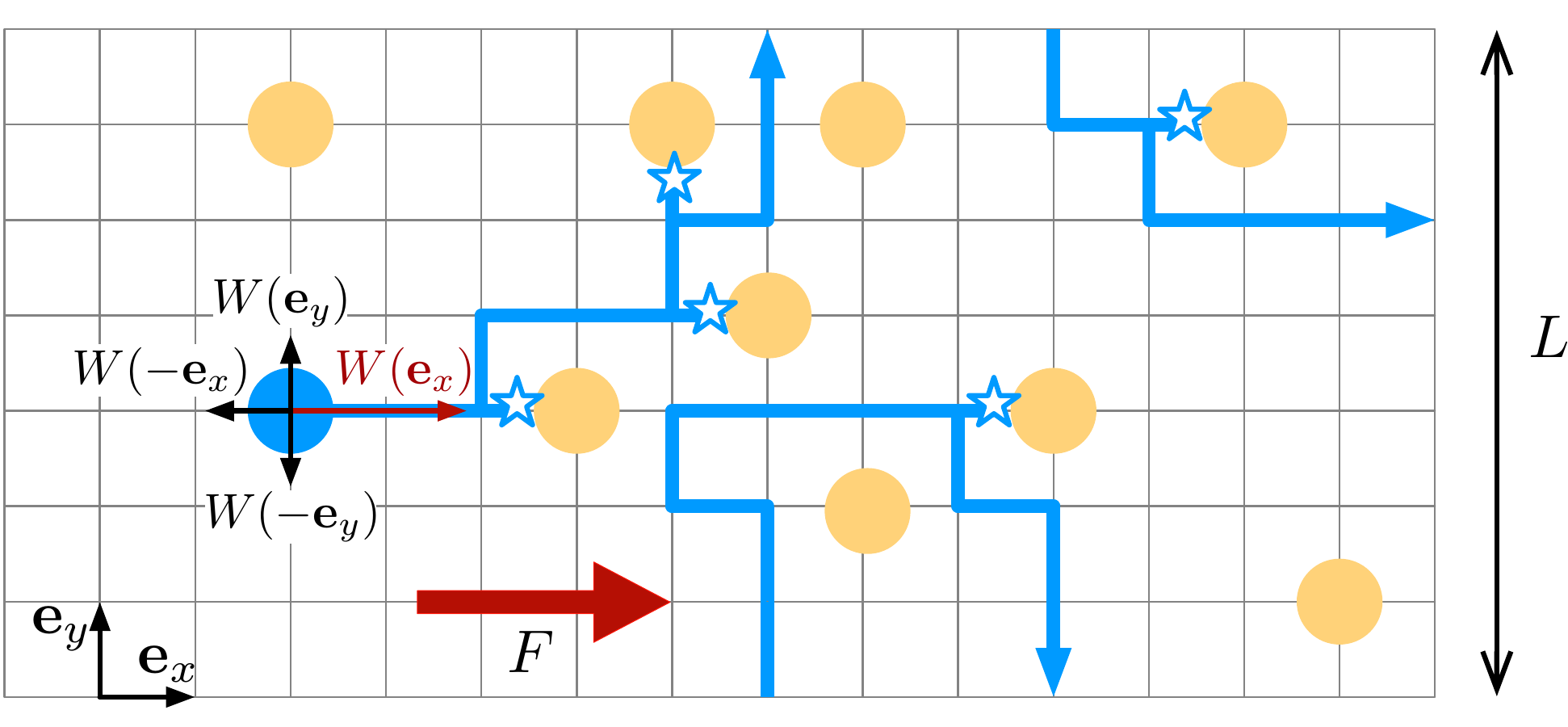}
\caption{Tracer moving on a lattice strip with identified edges (size $L$) under the action of a force $F$ along the strip axis. The tracer, scattered by quenched impurities (yellow circles), hops only on accessible sites; rejected transitions are indicated with a star.
}
\label{Fig_1_cylinder}
\end{figure}

\emph{Introduction.}---
Characterization of material properties by pulling  a mesoscopic tracer particle (TP) through a medium via optical or magnetic tweezers is at the essence of active-microrheology experiments \cite{Squires_2008, WP_2011, PV_2014}. Such a protocol has been applied to a large variety of systems including -- \emph{inter alia} -- colloidal suspensions \cite{PV_2014}, soft glassy materials \cite{Bocquet_2008,Bocquet_2012}, fluid interfaces \cite{Choi:2011gz}, and living cells \cite{HLRG_1999}, to mention a few. In contrast to passive microrheology \cite{Mason_Weitz_1995}, in which the thermally agitated motion of a TP is monitored, active microrheology in strong driving allows for the experimental exploration of a plethora of new phenomena in the full non-equilibrium regime such as force-thinning \cite{HSLW_2004, CB_2005, SMF_2010} and enhanced diffusivities \cite{WHVB_2012, WH_2013, Senbil_2019}. The scenario becomes even richer when crowding effects interplay with spatial confinement. In many experimental realizations the TP experiences also spatial constraints arising from space limitation due to boundaries \cite{Hanggi}. This is the typical situation observed when the TP moves into pores, narrow channels, or any other type of elongated quasi one-dimensional structure \cite{BIOSV_2018}.
\newline\indent
The dynamics in a complex environment turn out to be rather intriguing even at equilibrium since persistent correlations characterize the decay of correlation functions via slow power-law decays rather than exponential ones. In general, long-time tails and persistent memory effects are related to the singular behavior of transport coefficients \cite{AW_1970, ernst1971long, vB_1982}. For a $d$-dimensional fluid system the velocity autocorrelation function (VACF) encoding  time-dependent self-diffusion exhibits a decay of the form $t^{-d/2}$ due to slow diffusion of transverse momentum \cite{AW_1967, AW_1970, DorfmanCohen1970, Jeney2008, Franosch:2011df}. The above picture still persists, albeit with a different decay, when the dynamics takes place in a quenched disordered environment such as in the Lorentz model. The lack of momentum conservation and the repeated scatterings of the tracer with the same obstacle yield a long-time tail  $t^{-(d+2)/2}$ \cite{VANLEEUWEN1967457, WEIJLAND196835, EvB_1981, vB_1982, Hanna_1981,Mandal_2019}.  On the theoretical side, the Lorentz model has emerged as a paradigmatic model for the description of the dynamics in complex environment. In its lattice version, the TP explores a random array of fixed and impenetrable obstacles arranged on a lattice. The VACF of the two-dimensional Lattice Lorentz gas was calculated to first order in the density \cite{OK_nieuwenhuizen_3} and later confirmed by computer simulations \cite{FRENKEL1987385}. Progresses on the driven lattice Lorentz gas have been obtained in the last decade, in particular, the equilibrium dynamics to first order in the density \cite{OK_nieuwenhuizen_3, OK_nieuwenhuizen_2, OK_nieuwenhuizen_4, OK_nieuwenhuizen_1, VANLEEUWEN1967457, WEIJLAND196835, ernst1971long} has been generalized to an exact analytical solution for the case of a force pulling the tracer \cite{LF_2013, LF_2017, LF_2018new, LBF_2018, Squarcini_2024_LLG_discrete_time}.
\newline\indent
Over the years, several variations of the Lorentz Gas have been proposed, ranging from the Lorentz gas with fixed freely-rotating circular scatterers \cite{LLM_2003} to the many-particle one \cite{CEM_2009} which exhibits superdiffusive heat transport, to mention a few. Similarly, also the complementary situation of single-file diffusion in which a tracer particle moving in a dense environment of mobile particle on a lattice has been investigated analytically~\cite{Benichou_2013, Benichou_2014, Benichou_2016}. For single-file diffusion exact results are available~\cite{Illien_2013, Illien_2014}, as well as approximate ones \cite{Illien_2015} and numerical studies~\cite{Jack_2008, Basu_2014, BSV_2015}, including also the biased dynamics \cite{MB_1991} and the motion in an external potential \cite{BS_2009}. 
\newline\indent
Another closely related model that has been significantly investigated is the simple exclusion process~\cite{GM_2006, Mallick_2011}. Remarkable results are known for the large-deviation function for the tracer's displacement \cite{KMS_2014, IMS_2017} and for the dynamics with a moving defect \cite{Mallick_1996, LEM_2022}. While more recently the effects of random trapping have been studied for a tracer moving into a channel \cite{Shafir_2024}. The literature in this field is vast and comprises the driven one-dimensional lattice gases \cite{HMS_1999, KLMT_2022}. Driven diffusive systems in one dimension have been intensively studied in confining geometries such as narrow channels by means of mean field and simulations \cite{CMP_2017, MMP_2020}. Quite interestingly, the rate of particle overtaking can be used to tune the geometric confinement \cite{MM_2021}. Furthermore, it is known that these systems can exhibit interesting features such as absolute negative mobility \cite{CMP_2018} and bath-mediated interactions between two driven tracer particles moving in a narrow 2D channel \cite{MMP_2021}. In general, understanding how geometry, dimensionality and crowding affect the transport properties in crowded environments is a challenge for theory that this work wants to address, especially for the situation in which quenched disorder is present.
\newline\indent
In this Letter, we study and resolve the interplay of crowding, disorder, driving and spatial confinement by providing an exact solution for the quasi-confined lattice Lorentz model. The theoretical framework we employ allows us to obtain exact results to first order in the obstacle density for an arbitrary strength of the driving and confinement. Strictly speaking, the approach does not yield results at \emph{finite} density, which are far beyond our scopes. Moreover, at finite $n$, the channel will be obstructed with finite probability at some time $t_{*}$, which is large but nevertheless finite. In practice, $t_{*}$ is so large that the associated clogging phenomenon is never observed in our simulations. Technically, the quasi-confinement is introduced by wrapping the lattice onto a strip with identified boundaries such that it comprises $L$ parallel lanes with infinite extent along the axial direction. At time zero a step force $F\vartheta(t)$ is switched on and the tracer is pulled through the disordered lattice along the axial direction; see Fig.~\ref{Fig_1_cylinder}. One of the main result of this Letter is that -- already in equilibrium ($F=0$) -- the system exhibits a \emph{dimensional crossover} from two to one dimensions. This feature is demonstrated by the existence of two distinct long-time tails in the VACF characterized by different exponents, namely, a pre-asymptotic tail $t^{-2}$ followed by a slower tail $t^{-3/2}$ attained for long times. These two regimes are compatible with the theoretically predicted scaling $t^{-(d+2)/2}$~\cite{ernst1971long} provided the effective dimension is identified with $d=2$ at intermediate times and $d=1$ at long times.
\newline\indent
Beyond equilibrium properties, our analytical results allow us to explore also the regime of strong pulling where the linear-response regime breaks down. Our stochastic simulations reveal the domain of applicability of the theory to first order in the obstacle density. In particular, we show that the range of validity of the theory is density-dependent, i.e., the theory to first order in the density breaks down for large forces provided the density is small. Then, we provide exact analytical predictions for the velocity drift $v(t)$, the stationary diffusion coefficient and its dependence on the confinement. Our analytic results, exact to first order in the obstacle density, apply to an arbitrarily strong strength and confinement size $L$, encompassing the full dimensional crossover from the maximally confined periodic two-lane model ($L=2$) to the planar case, the latter is retrieved for $L \rightarrow \infty$.

\emph{Model.}--- We consider a tracer particle performing a random walk on a square lattice $\Lambda$ of unit lattice spacing, $\Lambda = \{ \mathbf{r} = (x,y): x = 1,\ldots, L_{x} , y=1,\ldots,L_y \}$ with $N = L_{x} L_{y}$ sites and periodic boundary conditions. The limit $L_{x} \to \infty$ is anticipated while the number of \emph{lanes} $L := L_{y}$ is kept fixed throughout; thus, the tracer hops on a strip with identified boundaries (length $L$), as sketched in Fig.~\ref{Fig_1_cylinder}.

The dynamics occur on a disordered lattice consisting of empty sites accessible to the tracer as well as randomly placed immobile hard obstacles of number density $n$ (fraction of excluded sites). Attempts of jumping onto an obstacles are rejected and the tracer remains at its initial position before the jump. The waiting time of the tracer at every site is exponentially distributed with mean waiting time $\tau$. With no loss of generality, we set $\tau =1$ which implies that time $t$ is measured in terms of the mean waiting time. The driving force $F$ is implemented by biasing the transition rates according to detailed balance, i.e. $W(\textbf{e}_{x})/W(-\textbf{e}_{x})=\exp(F)$, where $\textbf{e}_{x}$ is the unit vector along the $x$ direction and $F$ is measured in units of $k_{B}T$  \cite{HK_1987}. Upon normalization, $W(\pm\textbf{e}_{x})=\exp(\pm F/2)/[2+2\cosh(F/2)]$ while $W(\pm\textbf{e}_{y})=1/[2+2\cosh(F/2)]$, meaning that for large forces the transition parallel to the field is enhanced at the expenses of the rates in the directional perpendicular to the field.

\emph{Solution strategy.}---
In the absence of obstacles progress can be made on the analytical side by mapping the master equation for the site occupation probability to a Schr\"odinger equation~\cite{schutz,LF_2013}. The corresponding free Hamiltonian $\hat{H}_{0}$ describes the free motion on the empty lattice. Within this picture, the quenched disorder is accounted for by the Hamiltonian $\hat{H}_{0}+\hat{V}$ where the interaction potential $\hat{V}$ is chosen such that it eliminates any transition from and to the obstacles. The analytical solution of the dynamics to first order in the obstacle density relies on the scattering formalism borrowed from quantum mechanics~\cite{Ballentine}. Such an approach has been elaborated in the context of the Lorentz model in earlier works \cite{LF_2013,LF_2017, LF_2018new} and it constitutes the core of the analytical techniques underlying this Letter. The general ideas of the scattering formalism on a confined periodic lattice still apply to the case at hand, however, due to the confinement, the $x$ and $y$ directions are no longer interchangeable. This lack of symmetry brings an essential modification in the actual form of the scattering matrix for a single obstacle, which now requires the knowledge of lattice propagators on the strip. Details of the solution of the scattering problem for a single impurity together with extensive comparison with stochastic simulations will be presented elsewhere~\cite{JointPaper}.

\begin{figure}[htbp]
\centering
\includegraphics[width=\columnwidth]{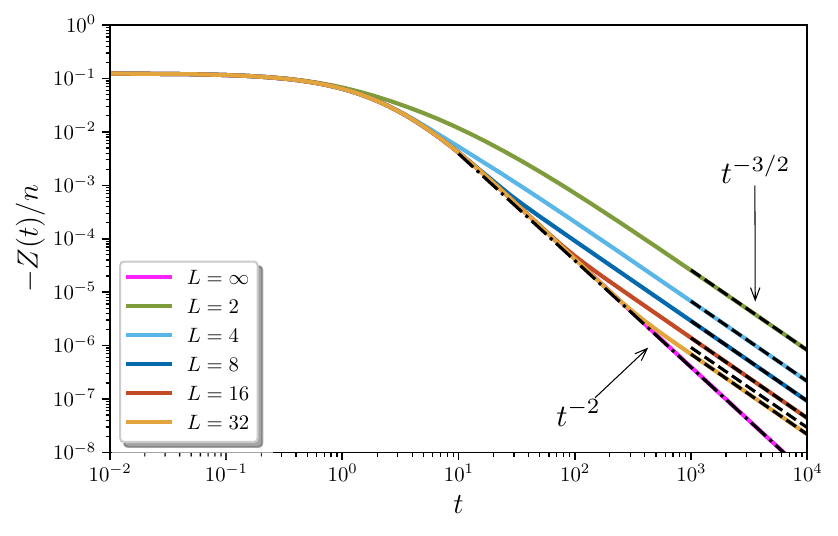}
\caption{The negative VACF for increasing confinement width~$L$. Dashed black lines correspond to the long-time tail asymptotically proportional to $t^{-3/2}$ [Eq.~(\ref{ltt_confined})]. The dot-dashed black line indicates the long-time tail $\sim (\pi/8) t^{-2}$ for the unconfined two-dimensional lattice Lorentz gas \cite{LF_2013}.}
\label{Fig_4_vacf}
\end{figure}

\emph{Discussion.}---
The occurrence of a dimensional crossover can be rationalized by examining the VACF, $Z(t)$, defined by
\begin{equation}
\label{ }
Z(t) := \frac{1}{2} \frac{\textrm{d}^{2}}{\textrm{d}t^{2}} \biggl[ \langle \Delta x(t)^{2} \rangle - \langle \Delta x(t) \rangle^{2} \biggr] \, .
\end{equation}
For sufficiently small driving the fluctuation-dissipation theorem (FDT) relates the mean drift velocity $v(t)$ along the unconfined direction to the VACF in equilibrium via
\begin{equation}
\label{FDT}
v(t) = F \int_{0}^{t}\textrm{d}t^{\prime} \, Z(t^{\prime}) \, ,
\end{equation}
where $v(t)$ is averaged over many realizations of the disorder. Passing to the frequency domain, the FDT reads
\begin{equation}
\label{15022024_1210}
\widehat{v}(s) = \frac{\widehat{Z}(s)}{s} F + O(F^{3}) \, ;
\end{equation}
while for the unconfined case the correction is stronger $O(F^3 \ln F)$. The hat in (\ref{15022024_1210}) denotes the Laplace transform, e.g., $\widehat{v}(s) : = \int_{0}^{\infty}\textrm{d}t \, v(t) \textrm{e}^{-st}$. The theory \add{\cite{LF_2018new}}, formulated in Laplace domain, yields an exact result for $\widehat{Z}(s)$,
\begin{equation}
\label{28092023_1600}
\widehat{Z}(s) = \frac{1}{4} + \frac{n}{4} - \frac{2n}{\Delta_{L}(s)} \, ,
\end{equation}
the above, as well as all our exact analytical results are shown only to first order in the obstacle density. The quantity $\Delta_{L}(s) = 4 - g_{00}(s) + g_{20}(s)$ is expressed in terms of the lattice Green's functions $g_{00}$ and $g_{20}$, the latter corresponds to a two-steps propagation along the $x$ direction while the former refers to zero steps. Long-time tails can be elaborated by analyzing the low-frequency expansion of lattice propagators; deferring to \cite{JointPaper}, we have $\Delta_{L}(s) = C_{L} + (8/L)\sqrt{s} + O(s)$ as $s \rightarrow 0$, with the confinement-dependent constant
\begin{equation}
\label{22092023_1208}
C_{L} = -4 + \frac{4}{L} \sum_{q=1}^{L-1} \sqrt{(2-\cos(2\pi q/L))^{2}-1} \, .
\end{equation}
For any finite $L$ the small-frequency behavior is dominated by the term $O(s^{-1/2})$ in the frequency-dependent diffusion coefficient $\widehat{D}(s)=\widehat{Z}(s)/s$ [Eq.~(\ref{28092023_1600})] which yields the long-time tail
\begin{equation}
\label{ltt_confined}
Z(t) \sim - \frac{8n}{\sqrt{\pi }LC_{L}^{2}} t^{-3/2} \, ,
\end{equation}
where $\sim$ denotes asymptotical equivalence for $t \rightarrow \infty$. Results for the entire time domain can be obtained by numerically inverting the Laplace transform, the result is shown in Fig.~\ref{Fig_4_vacf} for several values of $L$ including $L=\infty$ corresponding to the unconfined model.
The power law $t^{-3/2}$ in Eq.~(\ref{ltt_confined}) is characteristic for a one-dimensional system since it matches the long-time tail $t^{-(d+2)/2}$ for dimension $d=1$. This result can be interpreted by observing that for times much longer than a certain time scale $t_{L}$ the tracer has explored completely the periodic direction; hence, the effective one-dimensionality follows intuitively. However, for times much shorter than $t_{L}$ the tracer can't probe the finiteness of the confined direction and the decay is governed by the power law $t^{-2}$ corresponding to a two-dimensional system. The time scale describing the crossover between the two regimes is provided by the diffusive time $t_{L} \propto L^{2}$. It can be verified that for large $L$ and times $t = O(t_L)$ all curves in Fig.~\ref{Fig_4_vacf} asymptotically collapse onto a single master curve provided $t$ is rescaled in units of the crossover time $t_{L}$ \cite{JointPaper}.

\begin{figure}[htbp]
\centering
\includegraphics[width=\columnwidth]{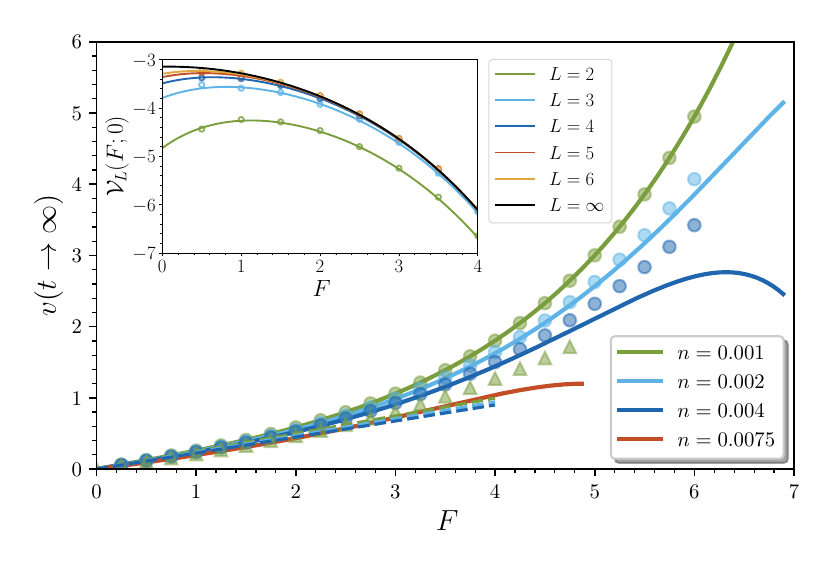}
\caption{Obstacle-induced velocity response in the stationary state $v(t \rightarrow \infty)$ as a function of the force $F$ for increasing obstacle density $n$ and system size $L$. Symbols correspond to stochastic simulations, solid lines represent the theory for arbitrary $F$, dashed lines indicate the linear response. The inset shows the function $\mathcal{V}_{L}(F;0)$.}
\label{Fig_2_vterminal}
\end{figure}

At long times a stationary state is reached, the latter is characterized by a terminal velocity $v(t \rightarrow \infty)$ that can be calculated analytically from the low-frequency behavior of the forward scattering matrix. Figure~\ref{Fig_2_vterminal} shows theoretical and simulation results for the terminal velocity as a function of the force. The terminal velocity then assumes the form
\begin{equation}
\label{v_freq}
v(t \rightarrow \infty) = v_{0} + n v_{0} + n v_{0} \mathcal{V}_{L}(F;0) \, ,
\end{equation}
where $v_{0}=(1/2)\sinh(F/2)$ is the bare velocity on the obstacle-free lattice and the function $\mathcal{V}_{L}(F;0)$, plotted in the inset of Fig.~\ref{Fig_2_vterminal}, encodes the correction due to the driving force, $F$, and finite size, $L$. The theoretical prediction [Eq.~(\ref{v_freq})] is compared to stochastic simulations in Fig.~\ref{Fig_2_vterminal} over a wide range of biases for different obstacle densities and system sizes. The expected breakdown of results to first order in $n$ can be inferred by considering how deviations become manifest for strong forces and high obstacle density.
It is interesting to consider also the regime of small forces and show how the theory yields exact predictions. For small driving and arbitrary finite size $L < \infty$ the terminal velocity can be expanded in the force as follows
\begin{equation} 
\label{26092023_1007}
v(t\rightarrow\infty) = \sum_{k=1}^{\infty} D_{L}^{(k)} F |F|^{k-1} \, ;
\end{equation}
which is non-analytic for $F=0$. Note that $v(-F) = - v(F)$ by reversing the force, as expected. The coefficients $D_{L}^{(k)}$ can be calculated in closed form and the leading term $k=1$ provides the linear response. For a small number of lanes:
\begin{equation}
\begin{aligned}
\label{19092023_1424}
D_{L=2}^{(1)} & = \frac{1}{4} - \frac{n}{4} (1+2\sqrt{2}) \\
D_{L=3}^{(1)} & = \frac{1}{4} - \frac{n}{8} (1+\sqrt{21}) \\
D_{L=4}^{(1)} & = \frac{1}{4} - \frac{n}{92} (-31+20 \sqrt{2}+12 \sqrt{3}+16 \sqrt{6}) \, .
\end{aligned}
\end{equation}
For small forces the linear behavior obtained by truncating Eq.~(\ref{26092023_1007}) at $k=1$ is tested in Fig.~\ref{Fig_2_vterminal} (dashed lines). The expression for $D_{L}^{(1)}$ becomes cumbersome as $L$ increases, nonetheless, for large $L$ we numerically established the asymptotic behavior $D_{L}^{(1)} = (1/4) - n [ (\pi-1)/4 + O(L^{-2}) ]$. Hence, the known result $D_{\infty}^{(1)} = [1 - n(\pi-1)]/4$ for the non-confined model \cite{LF_2013} is retrieved by taking $L \rightarrow \infty$. Beyond the linear-response term, the model on the unbounded plane ($L=\infty$) exhibits a non-analytic correction $\propto n F^{3} \ln|F|$ \cite{LF_2013}. Quite interestingly, as long as $L$ is finite the logarithmic term does not emerge, however, the confinement still yields a non-analytic behavior albeit now in terms of monomials of the form $F |F|^{k-1}$.

Let us discuss now the transport properties in the stationary state. For the unbiased model the late-time dynamics is characterized by the equilibrium diffusion coefficient
\begin{equation}
\label{20092023_1424}
D_{x}^{\rm eq}(t \rightarrow \infty) = \frac{1}{4} + n \left( \frac{1}{4} - \frac{2}{C_{L}} \right)
\end{equation}
to first order in the density $n$. By inspection of the VACF [Eq.~(\ref{ltt_confined})] it follows that the time-dependent diffusion coefficient relaxes towards its equilibrium value [Eq.~(\ref{20092023_1424})] as $t^{-1/2}$. The dependence on the system size $L$ enters via the confinement-dependent constant $C_{L}$, passing from $L=\infty$ to $L=2$ the bracket in Eq.~(\ref{20092023_1424}) decreases from $-(\pi-1)/4 \approx -0.54$ to $-(1+\sqrt{2})/2 \approx -1.21$, meaning that confinement enhances the reduction of the diffusion coefficient at equilibrium due to disorder. 

\emph{Summary and conclusions.}--- We have solved for the dynamics of a tracer particle performing a biased random motion in a disordered cylindrical lattice with driving force along the axis and with finite circumference mimicking the confinement. Results are valid, for any strength and confinement, to first order in $n$. We stress that at finite $n$ there is a finite probability that the disorder will obstruct the channel. This scenario -- although valid in principle -- is however not observed in the time range of our simulations; the latter suffices in order to extract quantities in the stationary state. Quite interestingly, even without an applied driving the system exhibits a dimensional crossover between two regimes of persistent anti-correlations characterized by different long-time tails. This finding is obtained by calculating the complete time dependence of the VACF to first order in the density $n$ of impurities. In particular, we obtain a long-time tail of the form $t^{-(d+2)/2}$ with effective dimension $d=1$ at infinite times, while at intermediate times $d=2$. These two regimes are separated by a temporal time scale $t_{L} \propto L^{2}$ which is interpreted as the time needed to explore the confined direction. 

We then showed that in response to a step force the non-equilibrium stationary state is characterized by a driving-dependent terminal velocity. Our analytical result for the terminal velocity is valid at both small and large driving force, $F$, and includes, as a particular case, the linear response regime. For the latter we find closed-form expressions for the mobility coefficients and their dependence on the confinement size $L$, showing thus how linear response is affected by geometrical constraints. Another striking fact emerging from the exact solution is that confinement alters the non-analytic dependence on $F$ of response functions. In this Letter, we examined this fact for the terminal velocity, which contains non-analytic terms of the form $F |F|^{k-1}$ while for the unconfined model the non-analyticities are of the form $F^{3} \ln |F|$. By comparing our analytical predictions with stochastic simulations, we found a good agreement and, in general, the latter improves by towards large values of the force provided the obstacle density is lowered. This feature shows how the analytical solution to first order in the density yields a non-uniform domain of validity of the theory itself, and simulations allow us to quantify this effect. When the external driving is removed the dynamics at infinite times is characterized by an equilibrium diffusion coefficient that we found analytically for any $L$. Our results show that the interplay of confinement and disordered effects yields a stronger suppression of the equilibrium diffusion coefficient for increasing obstacle density. 

From the analytic solution we learned that several features of the lattice Lorentz gas model are rather robust since they occur also in the presence of confinement. However, the actual form of non-analyticities in the response functions are very sensitive to confinement, as illustrated by the terminal velocity; this fact is further confirmed by analyzing the stationary diffusion coefficient \cite{JointPaper}. In addition, we expect our result to be transferable to three-dimensional systems. More precisely, the dimensional crossover we unveiled in this minimal model should occur also in quasi-1D pore-like or quasi-2D slab geometries in three dimensions, in such a context a richer phenomenology involving a hierarchical dimensional crossover is expected to occur. In our work we considered a quasi-confined system, however the presence of a restricted geometry can be implemented in many other ways. Nonetheless, we expect the picture to persist even when periodic boundary conditions are replaced by impenetrable walls. More on the speculative side, it would be very interesting to address in simulation studies the effects played by imperfect walls in which corrugation effects \cite{PMP} allow for slowly-varying cross sections and deposition of obstacles at surfaces.

\indent
\begin{acknowledgments}
\emph{Acknowledgments.}---
 This research was funded in
part by the Austrian Science Fund (FWF) 
through the Lise-Meitner Fellowship (Grant DOI 10.55776/M3300) 
 and 10.55776/P35673. TF gratefully acknowledges hospitality of Universit\'e Pierre et Marie Curie where parts of the project have been performed. 
\end{acknowledgments}


%

\end{document}